# High-Temperature Quantum Valley Hall Effect with Quantized Resistance and a Topological Switch

Ke Huang[1], Hailong Fu[1†], Kenji Watanabe[2], Takashi Taniguchi[3], Jun Zhu[1,4*]

**Affiliations**

[1]Department of Physics, The Pennsylvania State University, University Park, Pennsylvania 16802, USA.

[2]Research Center for Electronic and Optical Materials, National Institute for Materials Science, 1-1 Namiki, Tsukuba 305-0044, Japan

[3]Research Center for Materials Nanoarchitectonics, National Institute for Materials Science, 1-1 Namiki, Tsukuba 305-0044, Japan

[4]Center for 2-Dimensional and Layered Materials, The Pennsylvania State University, University Park, Pennsylvania 16802, USA.

*Correspondence to: jzhu@phys.psu.edu (J. Zhu)

†Present address: School of Physics, Zhejiang University, Hangzhou 310058, China

**Abstract**

Edge states of a topological insulator can be used to explore fundamental science emerging at the interface of low dimensionality and topology. Achieving a robust conductance quantization, however, has proven challenging for helical edge states. Here we show wide resistance plateaus in kink states – a manifestation of the quantum valley Hall effect in Bernal bilayer graphene – quantized to the predicted value at zero magnetic field. The plateau resistance has a very weak temperature dependence up to 50 Kelvin and is flat within a dc bias window of tens of mV. We demonstrate the electrical operation of a topology-controlled switch with an on/off ratio of 200. These results demonstrate the robustness and tunability of the kink states and its promise in constructing electron quantum optics devices.

**Main text**

Chiral and helical edge states arising from a non-trivial bulk topology are expected to be protected against backscattering and support a wealth of interesting topological, mesoscopic and interacting phenomena in low dimensions (*1, 2*). Chiral edge states of the quantum Hall (QH) and quantum anomalous Hall (QAH) effects are ballistic over long distances (*1, 3-6*), harbor the physics of a chiral Luttinger liquid (*7*), and are instrumental to the understanding of fractional and non-Abelian braiding statistics (*8, 9*). Helical edge states of the quantum spin Hall (QSH) and quantum valley Hall (QVH) effects occur at zero external and internal magnetic field, which is generally more compatible with device applications (*10-28*). Previously, QVH internal edge states, known as the kink states, were realized in Bernal bilayer graphene (BLG) using lithographically patterned gates, and operations of a valley valve and electron beam splitter were demonstrated (*22, 23*). The tunable beam splitter is analogous to the action of a quantum point contact, which is a powerful element used in a wide range of fundamental research and quantum devices (*29*). The absence of precise resistance quantization, which implies backscattering is still present, is a critical impediment to the pursuit of some of the most exciting prospects of helical 1D systems, such as topological superconductivity (*27, 28, 30, 31*), helical Luttinger liquid physics (*13, 32-34*) and the development of edge-based quantum transport devices (*11, 18, 23, 28, 35*).

In this study, we report on the precise quantization of QVH kink state resistance to better than 1% of the expected $R = h/4e^2 = 6453$ Ω ($h$ is the Planck's constant and $e$ is the elementary electron charge) at zero magnetic field and its persistence to tens of Kelvin. We further demonstrate the fast and repeatable operation of a topological switch, the mechanism of which is based on an electric-field-controlled topological phase transition unique to the kink states. The ballistic transport, in situ programmability, and scalable fabrication of the kink states lay a strong foundation for future studies exploring the fundamental science and application potentials of this helical 1D system.

**Device structure and characterization**

Figures 1A and B illustrate the topological origin of the QVH effect in BLG (*16-20*) and a dual-split-gate-based experimental scheme respectively (*16, 22-24, 26*). The four split gates, i.e. the left/right top/bottom gates (L/R T/B Gs) define two independently gated regions. We induce bulk gaps of opposite sign in the two regions and the resulting band inversion at the nanoribbon junction creates valley-momentum locked kink states inside the junction. The device structure in Fig. 1B enables an electrical control of the transition between the topological phase with the existence of the kink states, and the trivial phase without it. This situation is illustrated in Fig. 1C and underlies the working mechanism of a topological switch (*11*), which we will demonstrate. Each valley (K and K′) contains four chiral modes. Ballistic transport of the kink states results in a two-terminal junction resistance of $h/4e^2 = 6453$ Ω. Electron backscattering requires an intervalley scattering that involves a large momentum transfer of order (lattice constant)$^{-1}$; such events are highly suppressed in highly crystalline BLG but not completely absent because of residual Coulomb impurities. Indeed, the mean free path of the kink states is at most a few μm in previous experiments and the zero-field resistance is at least hundreds of Ω away from $h/4e^2$ with no clear quantization plateaus (*21-24, 26*).

The key technical advance of this work is the incorporation of a graphite/hexagonal boron nitride (h-BN) stack as a global gate (GG) to tune the Fermi level of the kink states. Previous devices in Refs. (*22, 23*) used a doped Si/SiO$_2$ stack, which was shown to be an important source of potential disorder (*36*). We use dry van der Waals transfer, annealing, and reactive ion etching to create the graphite split LBG and RBG, an atomic force microscope (AFM) image of which is shown in Fig. 1E. Additional transfers and e-beam lithography are used to complete the device (Figs. S1-3) (*37*). Figure 1D shows a false-color scanning electron microscope (SEM) image of K23 with a 61 nm-wide, 400 nm-long junction. The two-terminal resistance of the junction $R = R_k + R_c$ includes both the kink state resistance $R_k$ and the contact resistance $R_c$; $R_c$ is accurately determined using the resistance quantization of the quantum Hall effect (Fig. 2C) and is typically a few hundred Ω.

We characterize the five gates (LTG, LBG, RTG, RBG, and GG) (See Fig. S4) and place the dual gated areas (shaded gold in Fig. 1D) in the middle of a *D*-field induced bulk gap *Δ*, with a typical size of tens of meV. Figure 1F shows the evolution of the junction resistance $R$ in device K22 as the device is tuned from the "− −" gating configuration (both left and right *D*-fields pointing down) to the "− +" configuration (left down, right up). $R$ reaches MΩ in the "− −" configuration, where the residual conductance comes from localized states in the nanoribbon junction (*22, 23*). It saturates to 6 - 7 kΩ in the "− +" configuration, confirming the presence of the kink states in the topological regime.

**Resistance quantization at $B = 0$ T**

We apply a voltage to the GG, $V_{GG}$, to tune the Fermi level $E_F$ inside the junction. Figure 2A plots an example of the junction resistance $R(V_{GG})$ in device K22. Inside the gap where only the kink states are present, $R$ exhibits a flat resistance plateau of $R = (6557\pm12)$ Ω. Similar plateaus, with $R$ very close to expected $h/4e^2$, are observed at other $D$-fields (See Fig. S5 for a systematic investigation of $R(V_{GG})$ under different $D$-fields). Excellent resistance quantization is also observed in other devices. Figure 2B plots a few exemplary $R(V_{GG})$ traces in K24 while measurements in WK05 are given in Fig. S7.

The breaking of time-reversal symmetry through the application of a magnetic field does not destroy the kink states (*19, 22, 23*). Indeed, our measurements show that $R$ varies by 53 Ω from 0 to 9 T, likely owing to small changes in $R_c$ (Fig. S6B). As the $B$-field increases, the conduction and valence bands of the nanoribbon junction evolve into Landau levels (LLs), with each LL contributing $4e^2/h$ to the total conductance. Thus, the total junction resistance is

$$R = R_c + \frac{h}{4e^2(1+n)} \text{ (Eq. 1)}$$

where n = 0, 1, 2… represents the kink only (n = 0) and the number of occupied LLs. The main panel of Fig. 2C shows the measured $R(V_{GG})$ at $B = 0$, 1, and 2 T, with the LL quantization already well developed at $B = 1$ T. The measured $R$ values are extremely well described by Eq. 1 and a linear fit to the $B = 1$ T data (Fig. 2D) yields $R_c = 114$ Ω and $R_k = R − R_c = 6443$ Ω. $R_k$ deviates from the expected value of $h/4e^2 = 6453$ Ω by 10 Ω. Performing similar analysis at other $B$-fields and temperatures, we find the resulting $R_k$ to always fall within 0.4% of $h/4e^2$ (Table S3). We have also examined the variations of the kink state plateau resistance at different $D$- and $B$-fields in multiple devices (Figs. S5-S7). These analyses allowed us to place 1% as an upper bound to the accuracy of $R_k$ in our experiment. Our results unambiguously demonstrate the precise and robust quantization of the kink state resistance. Quantization to this level has so far only been observed in chiral edge states (*1, 3, 5*), whereas helical edge states in any materials are more vulnerable to backscattering. The excellent ballisticity and scalability of the QVH kink states make it an attractive platform to construct electronic waveguides and explore quantum information concepts such as a flying qubit (*38*).

**The temperature dependence**

The temperature dependence of the kink states is examined in Fig. 3. Figure 3A plots exemplary $T$-dependent $R(V_{GG})$ traces taken in device K22. The junction resistance $R$ remains constant at low temperatures and decreases slowly with increasing $T$. Figure 3B plots the normalized plateau center resistance $R$ vs $T$ at selected bulk $D$-fields in K22 and WK05. $R(T)$ has very weak $T$-dependence, especially when $D$, hence the bulk gap $\varDelta$, is large. Figure 3C plots the temperature $T_0$ at which $R$ deviates from its low-$T$ value by 1, 2, 3, and 5%. Notably, at $D = 450$ mV/nm or $\varDelta \sim 58$ meV (*39*), $T_0 = 12$ K for 1% and 46 K for 3%. This high level of $T$-independence is a hallmark of topological edge states but has only been observed previously in the QH effect, whereas QAH and QSH systems appear to be more vulnerable to temperature (*3-6, 10, 12, 14, 40, 41*).

The complex behavior of $R(T)$ shown in Fig. 3B can in fact be well explained by recognizing the fact that the gapped BLG bulk conducts in parallel to the kink states, as the inset illustrates. In this temperature range, the $T$-dependence of gapped BLG is well understood as consisting of thermally activated band transport and nearest neighbor hopping (NNH) inside the gap, i.e.

$$R_{\text{bulk}}^{-1} = R_1^{-1} e^{(-\Delta/2k_BT)} + R_2^{-1} e^{(-E_2/k_BT)} \text{ (Eq. 2)}$$

where $R_1$ and $R_2$ are constants, $E_2$ is the average energy difference between adjacent impurity sites, and $k_B$ is the Boltzmann constant (*39, 42-45*). Figure 3D plots the simulated $R_{\text{bulk}}(T)$ using $R_1 = 1$ k$\Omega$, $R_2 = 200$ k$\Omega$, $E_2 = 1.2$ meV and $\Delta$ [meV] = -5.7 + 0.13$D$ [mV/nm] (See Section 6 of (*37*)). The simulated total junction resistance $R = R_{2K}//R_{\text{bulk}}$ is plotted in Fig. 3B as dashed lines and is in excellent agreement with the measured data. The fact that this simple model appears to have captured all of the complex features of the $R(T)$ traces, qualitatively and quantitatively, gives us confidence in its validity.

The above analysis suggests that the intrinsic $T$-dependence of the kink states remains undiscernible, i.e., phonon-assisted intervalley scattering is negligible, up to 50 K. We note that BLG phonons carrying the required intervalley momentum transfer $\delta k$ of the 1D kink states possess a minimum energy $\hbar \omega_{min}$ as high as 68 meV when the junction aligns with the zigzag crystallographic orientation of the BLG (*46, 47*). When the junction deviates from the zigzag direction, $\delta k$ is reduced but remains large (See Section 3 of (*37*) for an estimate of $\delta k$ in our devices). Thus, the kink states may be immune to phonon-assisted backscattering to temperatures on order of 100 K. The further suppression of bulk conduction, e.g. through increasing $\Delta$ and reducing impurity sites, will enable us to fully examine its potential.

**The dc bias dependence**

Next, we examine the dc bias dependence of the kink states. Figure 4A shows a false-color map of differential conductance $G(V_{dc}, V_{GG})$ in device K22. Figure 4A is reminiscent of the transport spectroscopy of an insulator, only here $G = 4e^2/h$ inside the diamond-like region. Figure 4B plots $G(V_{dc})$ traces taken in WK05 with $V_{GG}$ positioned at midgap and at selected $D$-fields. $G(V_{dc})$ is nearly flat at low biases and increases sharply at a bias threshold $V_{dc}$, which increases with increasing $D$. $V_{dc}(D)$ determined from measurements is plotted in Fig. 4C. It exhibits an approximate linear dependence with a slope of 0.09 mV/(mV/nm) and reaches several tens of mV at large $D$. Our data can be well understood using the band diagram shown as the inset of Fig. 4C, where the sharp onset of $G$ at $V_{dc}$ corresponds to the electrochemical potential of the electrodes reaching the band edge of the nanoribbon junction. We obtain $\Delta'$ [meV] = -4 + 0.09$D$ [mV/nm] from data, which is consistent with band structure simulations of a finite-width nanoribbon junction (*22*). The behavior of the kink states at finite dc biases attests to the cleanness of the gapped region and further demonstrates the robust quantization of the kink states. The large bias window of several tens of mV facilitates potential single-electron manipulations.

**A topological switch**

As illustrated in Fig. 1C, the kink state possesses a unique property, that is, its presence can be turned on and off using an electric-field-controlled topological to trivial insulator transition. This topological phase transition-driven switch is fundamentally different from that of a conventional transistor, which operates through carrier concentration change. In the literature, topological switches based on phase transitions were proposed for quantum spin Hall and topological crystalline insulators (*11, 48*) but these proposals are difficult to realize experimentally. Figure 5 shows the operations of our kink state topological switch, where the junction conductance is switched between ~ $4e^2/h$ and ~ 0 repeatedly as the gating configurations are alternated between the "+ −" and "− −" configurations through the voltage swing of a few volts. This operation is very reproducible. The on/off ratio is approximately 200, which can be further increased by reducing the "off" state conductance. The rise time of ~ 6 ms is limited by the charging speed of the gates. Neither parameter is limited by the intrinsic properties of the kink states and both have much room for further improvement.

**Discussion and outlook**

In this work, we have achieved the quantization of the QVH effect at zero magnetic field and to high temperatures of tens of Kelvin. Its manifestation in bilayer graphene, the kink state, distinguishes itself from other chiral and helical edge state systems in its electrical creation and wide in situ tunability, which makes it an attractive platform to construct on-chip quantum electronics that mimic the functionalities of quantum optics. Clean material and device construction are crucial to the realization of this potential. Significant improvement was made in this work, which led to excellent ballistic charge transport of the kink states and resistance quantization. This quantization is also robust in wide temperature and voltage ranges owing to relatively few in-gap states and the high energies of phonons in graphene. Our experiments point to lowering defect and charged impurity states as a vital path to achieving quantized transport, especially for helical edge states, where the wavefunctions of the time-reversed pair co-locate spatially. We envision the construction of a quantum interconnect network using the kink states as the backbone and integrating the functional elements demonstrated to date, namely switch, waveguide, valve, and beam splitter (*23*). Such network may be used to carry quantum information on-chip over a long distance, for which the preservation of quantum coherence and entanglement is essential. The demonstration of long spin relaxation length in bilayer graphene (*49*) bodes well for this vision and phase coherent transport will be explored in future studies (*38*). Kink states are also excellent candidates to explore the physics of 1D helical Luttinger liquid through tunneling experiments analogous to measurements performed for chiral edge states at a quantum point contact (*7, 50*). The helical nature of the edge states is expected to give rise to different critical behaviors (*33*). Coupling kink states to superconductivity is another interesting direction to pursue and recent experiments using domain wall modes in twisted bilayer graphene have yielded promising initial results (*51*).

**Acknowledgment**

We thank Fan Zhang, Kaijie Yang, Yuval Oreg, and Noam Schiller for the helpful discussions and Ruoxi Zhang for the LabVIEW program used in measurements. **Funding:** K. H. and J. Z. are supported by the National Science Foundation through grants NSF/DMR-1904986 and the Department of Energy through grant DE-SC0022947. DE-SC0022947 supported a portion of the measurements and data analyses. H.F. acknowledges the support of the Penn State Eberly Research Fellowship and the Kaufman New Initiative research Grant No. KA2018-98553 of the Pittsburgh Foundation. K.W. and T.T. acknowledge support from the JSPS KAKENHI (Grant Numbers 20H00354, 21H05233 and 23H02052) and World Premier International Research Center Initiative (WPI), MEXT, Japan. Work performed at the National High Magnetic Field Laboratory was supported by the NSF through NSF/DMR-2128556 and the State of Florida.

**Author contribution:** K. H. and J. Z. designed the experiment. K. H. fabricated the devices and made the measurements. H. F. assisted in device fabrication and measurement. K. H. and J. Z. analyzed data. K. W. and T. T. synthesized the h-BN crystals. K. H. and J. Z. wrote the manuscript with input from all authors.

**Competing interests**: The authors declare no competing interests.

**Data and materials availability:** The data that support the plots within this paper are available from Harvard Dataverse (*52*).

**Supplementary Materials**

Figs. S1 to S9

Tables S1 - S3

References (53-54)

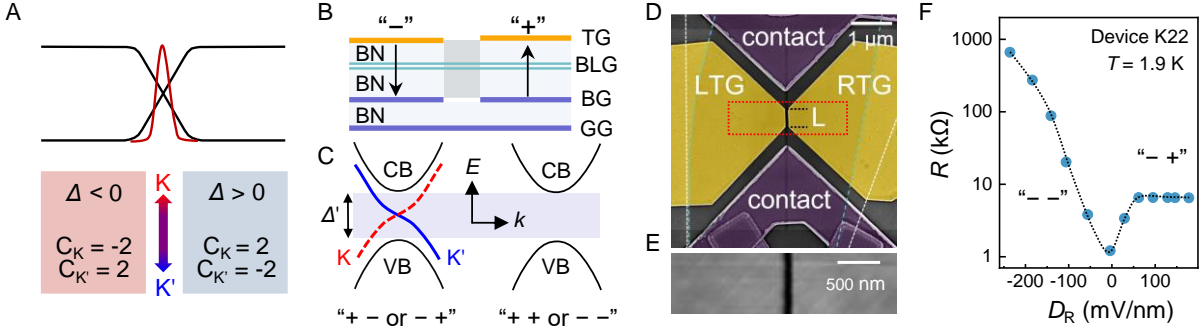

**Fig. 1. The quantum valley Hall kink states in bilayer graphene.** (**A**) The schematic band diagram showing the inverted bands, the wave function of the kink states and its topological origin. The valley Chern number changes by +4/−4 in the K/K′ valley across the junction, giving rise to four chiral modes in each valley. (**B**) illustrates the schematic side view of our device structure. The $D$-field points up in a "+" configuration. (**C**) One-dimensional band diagram of the junction (gray area in (B)) in the "+ − / − +" (topological) or "+ + / − −" (trivial) gating configurations. C CB: conduction band, VB: valence band. (**D**) A false-color SEM image of device K23. The Ti/Au top split gates align with the graphite bottom split gates. The BLG sheet is outlined in cyan and the global graphite gate is outlined in white. The side contacts are colored purple. They contact the kink states through the heavily doped access region at the two ends. The junction length $L$ = 400 nm in all our devices. (**E**) An AFM image of the 61 nm-wide bottom graphite split gates in the area enclosed by the red box in (**D**). The surface is clean after annealing. In device K22, K24 and WK05, the width is 74, 71 and 74 nm respectively. (**F**) The junction resistance $R$ in device K22 as a function of $D_R$. $D_L$ = -161 mV/nm. $B$ = 0 T. $T$ = 1.9 K. The data are extracted from Fig. S5B. See (37) for the fabrication and characterization of the devices.

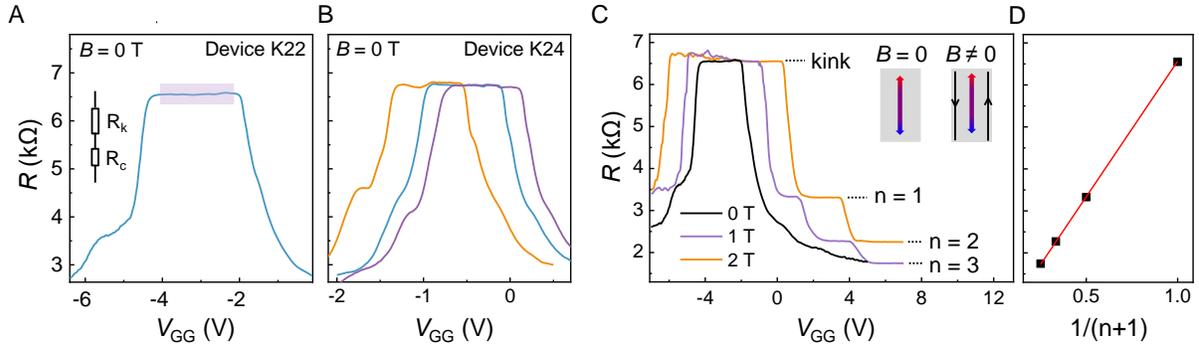

**Fig. 2. The resistance quantization of the kink states.** (**A**) Junction resistance $R$ vs $V_{GG}$ in device K22. In the purple shaded region, $R = 6557 \pm 12\ \Omega$. (**B**) plots $R(V_{GG})$ at selected $D_L$ values in K24. $D_L$ = -296 (orange), -199 (blue), and -150 (violet) mV/nm. $D_R = 241$ mV/nm for all. The traces are shifted in $V_{GG}$ owing to misalignment on the left gates. $B = 0$ T for both (A) and (B). (**C**) $R(V_{GG})$ at selected magnetic fields as labeled in the graph. The inset illustrates the appearance of QH edge states in the nanoribbon junction in a finite $B$-field. The index n marks the number of occupied Landau levels. (**D**) $R$ vs $1/(n+1)$. Points are taken from the data at 1 T in (C). A linear fit (red line) using Eq.1, yields $R_c = 114\ \Omega$ and $R_k = 6443\ \Omega$. (A), (C) and (D) are from device K22 with $D_L$ = -161 mV/nm and $D_R = 146$ mV/nm. $T = 1.9$ K for all data in this figure.

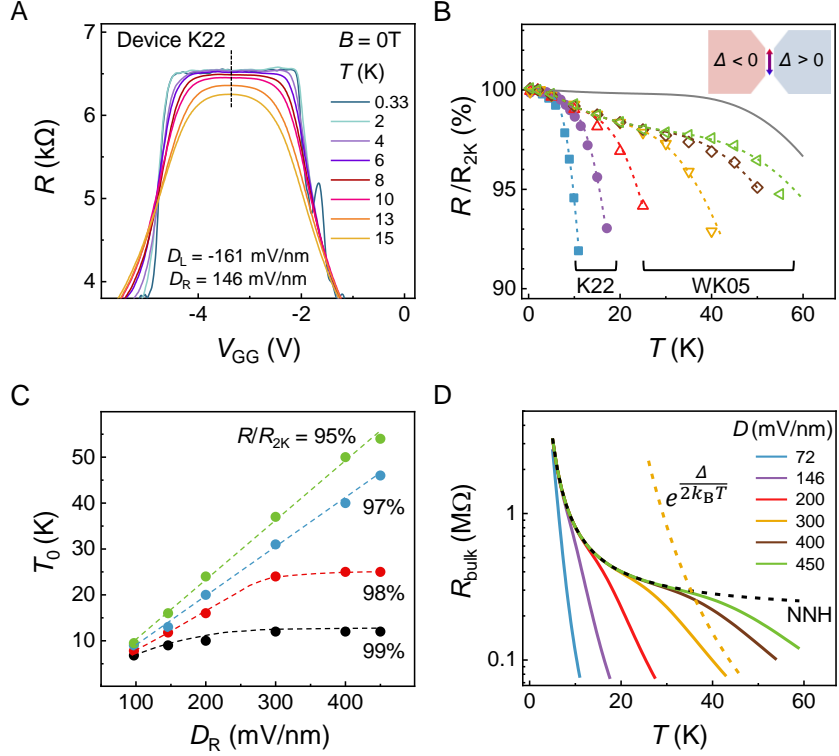

**Fig. 3. The temperature dependence of the junction resistance.** (**A**) $R(V_{GG})$ measured in device K22 at temperatures from 0.33 K to 15 K as labeled in the graph. (**B**) Normalized plateau center resistance $R/R(T = 2\text{ K})$ vs $T$ at selected $D$'s. Solid symbols are from device K22, where $D_L = -161$ mV/nm is fixed while $D_R$ changes in this device. Open symbols are from device WK05, where $D_L = -D_R$. The $D_R$'s of both solid and open symbols follow the legend in (D). Raw data in WK05 are shown in Fig. S7. The dashed lines plot $R = R_{2K}R_{bulk}(T)/(R_{2K} + R_{bulk}(T))$, where $R_{bulk}(T)$ is given by Eq. 2 and plotted in (D). The inset illustrates the parallel bulk and kink state contributions to $R$. $B = 0$ T for all data in this Figure. (**C**) plots the temperatures $T_0$ at which $R/R_{2K} = 95\%$, 97%, 98% and 99%, using measurements in (B). Dashed lines are guide to the eye. (**D**) Simulated bulk resistance $R_{bulk}(T)$ using Eq. 2 for different $D$-fields as labeled in the graph. The orange and black dashed lines plot the thermally activated and the NNH terms separately for the orange trace. The two traces cross at 35 K. The total $R_{bulk}(T)$ is dominated by thermal activation above 35 K and NNH below 35 K. The expanding dominance of NNH at large $D$ leads to a saturated $R_{bulk}$ and a shoulder-like feature in $R/R_{2K}$ in (B). The gray solid line in (B) plots a simulated scenario where the NNH resistance in the green dashed line is increased by ten-fold, via, e.g. reducing the number of impurity states or reducing the width of the gapped bulk. In this scenario, the contribution of $R_{bulk}$ becomes negligible below 40 K.

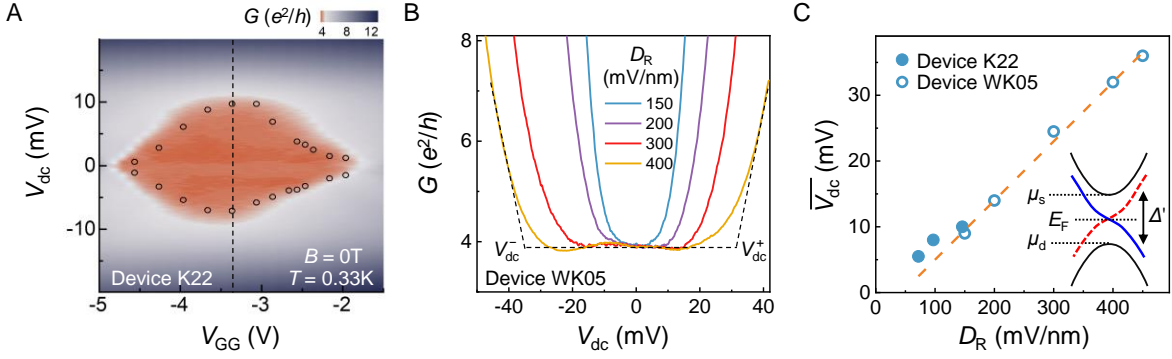

**Fig. 4. The dc bias dependence of the kink states.** (**A**) A false-color map of differential conductance $G(V_{dc}, V_{GG})$ in device K22. $D_L = -161$ mV/nm and $D_R = 146$ mV/nm. The onset of high conduction, defined as $G = 4.3e^2/h$ (white color in the map) traces out a diamond-like shape. (**B**) $G(V_{dc})$ sweeps obtained in device WK05 with $E_F$ positioned at midgap and at selected $D$'s as labeled. $D_L = -D_R$. The black dashed lines illustrate a second way of defining the conduction onset dc bias $V_{dc}$. Open circles in (A) are obtained using this method. (**C**) The average onset bias $\overline{V_{dc}} = (V_{dc}^+ - V_{dc}^-)/2$ as a function of $D_R$ in both devices. $D_L = -161$ mV/nm is fixed in K22. $D_L = -D_R$ in WK05. The inset illustrates the onset of band conduction at a finite dc bias $eV_{dc} = \mu_s - \mu_d = \Delta'$, where $\Delta'$ is the band gap of the nanoribbon junction. The orange dashed line has a slope of 0.09 mV/ (mV/nm). $B = 0$ T and $T = 0.33$ K for all data in this figure

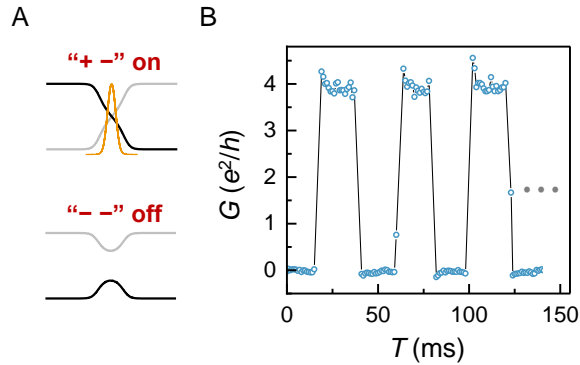

**Fig. 5. A topological phase transition-controlled switch.** (**A**) Band diagrams for the "on" (upper panel) and "off" states (lower panel) of the switch using calculations in Ref. (*22*). (**B**) The operation of the switch for a few cycles. $D_L = 57$ mV/nm ($V_{LTG} = -1$ V, $V_{LBG} = 0.467$ V) and -75 mV/nm ($V_{LTG} = 1$ V, $V_{LBG} = -0.687$ V) respectively for the on/off state. $D_R$ is fixed at -170 mV/nm. The on/off ratio is about 200. $B = 0$ T and $T = 0.33$ K. Fig. S9 provides more examples of the switching operations of the junction.

# Supplementary Materials for

## High-Temperature Quantum Valley Hall Effect with Quantized Resistance and a Topological Switch

Ke Huang *et al.*

Corresponding author: Jun Zhu, jzhu@phys.psu.edu (J. Zhu)

**The PDF file includes:**

1. **Device fabrication**

We combine van der Waals dry transfer, Ar/O$_2$ annealing, and precision lithography introduced in Ref. (*22*, *23*, *53*) to fabricate kink state devices. We first assemble and transfer the h-BN/global graphite gate stack to a SiO$_2$/doped Si wafer. The stack is then annealed in Ar/O$_2$ at 450 °C, followed by the transfer of the bottom graphite gate sheet. This process is described in detail in our previous work (*53*). We then use e-beam lithography and reactive ion etch (RIE) (O$_2$ plasma) to pattern the bottom split gates with a narrow split of typically ~ 70 nm wide using polymethyl methacrylate (PMMA) as the resist. PMMA is developed in methyl isobutyl ketone (MIBK)/isopropyl alcohol (IPA) 1:1 at ~ 4 °C which provides a better e-beam lithography resolution. Subsequent annealing in Ar/O$_2$ removes the PMMA residue completely, leaving a pristine surface behind. An optical and AFM image of K22 at the completion of this step is shown in Figs. S1A and B respectively. Graphite trenches used for aligning the top and bottom split gates are also made concomitantly and shown in Figs. S1A and S2A. We then assemble and transfer a h-BN/bilayer graphene/h-BN stack using conventional dry transfer. The final stack is shown in Fig. S1C. Finally, we make electrical contacts and deposit the top split gates using methyl methacrylate (MMA)/PMMA bilayer and cold development. An SEM image of a finished device is shown in Fig. 1D of the main text.

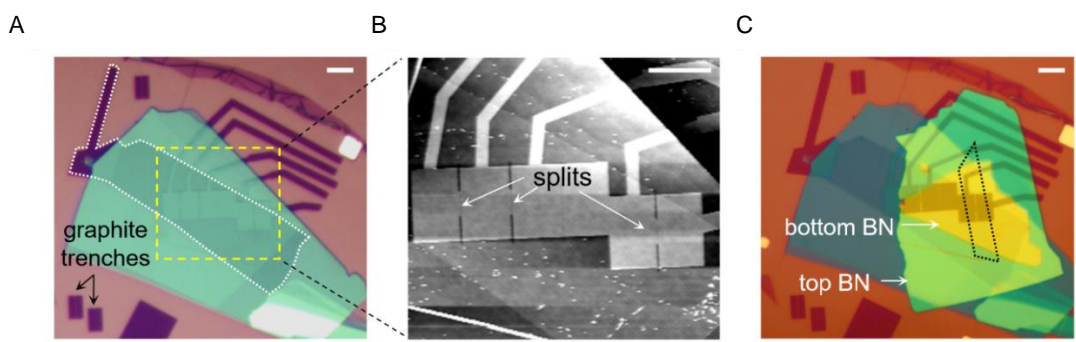

**Figure S1: Device fabrication. A**, An optical image of the K22 stack including the etched graphite bottom split gates and trenches made for alignment. The white dashed line outlines the graphite global gate (GG). **B**, AFM image of the area inside the yellow box in A. In areas of interest, the surface is clean of polymer residue. The widths of the splits vary from 74 to 90 nm, as measured by AFM. **C**, An optical image of K22 after the transfer of the h-BN/BLG/h-BN stack. The black dashed line outlines the BLG. All scale bars are 5 μm.

## 2. The alignment of the top and bottom gates

One of the essential steps of the fabrication is to align the top and bottom split gates, which largely follow the procedures described in Ref. (*22*) with the SEM imaging step now replaced by AFM to avoid potential damage/charging to the h-BN dielectric layer. A sharp tip (Budget Senors SHR300) with a typical tip radius of 1nm is used to obtain higher resolution in the AFM imaging. Briefly, dummy graphite trenches labeled as graphite in Fig. S2 were concomitantly etched and spatially registered to the bottom split gates of the real device. Prior to the patterning of the top gates, we perform the alignment procedure which includes the following steps. 1. Pattern a set of metal trenches designed to center on the dummy graphite trenches as shown in Fig. S2. During the same step, a new set of e-beam alignment markers registered to the metal trenches are also patterned. 2. Image the graphite and metal trenches using AFM to determine the relative shift between the two (Fig. S2B and C). 3. Correct the design of the top split gates with this shift and pattern them using the e-beam alignment markers freshly produced in step 2. The e-beam writer we use is Vistec EBPG 5200 with an overlay accuracy ≤ 5 nm.

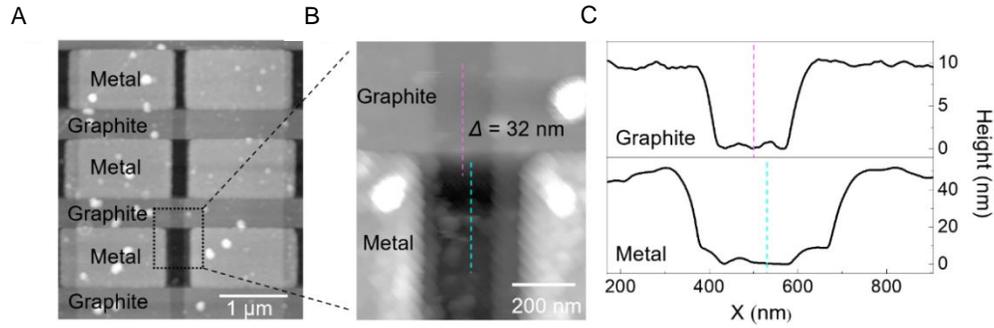

**Figure S2: Calibrating the misalignment A**, AFM image of the graphite and metal trenches. **B**, An enlarged image of the boxed area in A with the center locations marked by dashed lines determined from line scans shown in **C.** From device K23.

## 3. Phonon-assisted intervalley scattering

In this section, we provide an estimate for the impact of phonon-assisted intervalley scattering. Treating the kink state junction as a 1D graphene nanoribbon (GNR), we follow the zone-folding technique illustrated in Fig. S3A (*46*) to project the 2D Brillouin Zone (BZ) of the BLG to the 1D BZ of the junction.

When the junction is oriented along an angle θ relative to the zigzag crystallographic edge of the BLG sheet (θ is bound between 0 and 30° since zigzag edges repeat every 60°), the 1D BZ intervalley distance $\delta k = \frac{8\pi}{3a} \cos(\theta + 60°)$ nm$^{-1}$ (*46*), where a = 0.246 nm is the lattice constant of bilayer graphene. Specifically, $\delta k$ =17 nm$^{-1}$ or 0 for θ = 0 (zigzag) or 30° (armchair) respectively.

We estimate the orientation of the kink state junction in devices K22 and WK05 using AFM images as shown in Figs. S3B and C. Our BLG sheets contain sharp edges that are 30º, 60º, 90º, 120º, and 150º, with respect to one another. We associate them with the zigzag or armchair edges of the crystal (*54*). We avoid patterning junctions along the direction of any sharp edges because we cannot distinguish the armchair edge from the zigzag edge by optical microscopy. Fig. S3B shows that the junction in K22 is approximately 6º off from a sharp edge, i.e. $\theta = 6/24º$ if the edge is zigzag/armchair. This results in an intervalley distance of $\delta k = 13.8$ or $3.5$ nm$^{-1}$ in K22. Similarly, $\delta k = 12.2$ or $5.3$ nm$^{-1}$ in WK05. The lowest energy phonons in BLG that can provide such large momentum transfer to scatter an electron from the K to the K' valley of the kink states are the ZA phonons (*47*). Table S1 shows their corresponding energies. It is clear that so long as the junction is not too close to the armchair orientation, phonon-assisted backscattering only activates at high temperatures. This offers the kink states a potentially very wide temperature operation range, up to order 100 K.

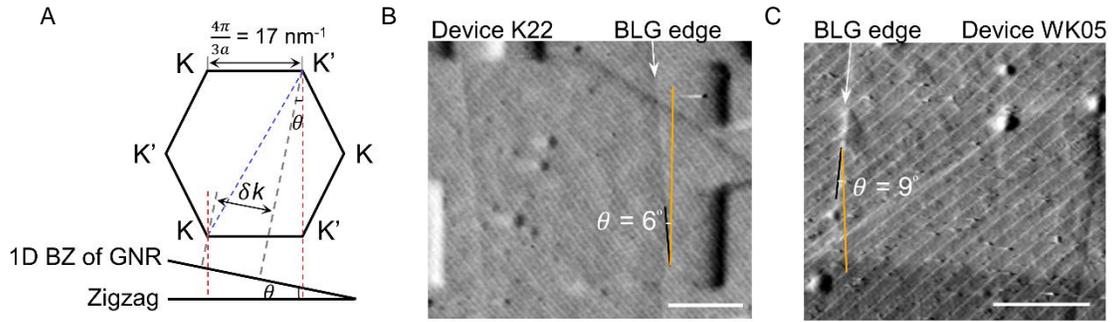

**Figure S3: Intervalley momentum transfer in a kink state junction.** **A**, A schematic diagram illustrating how to calculate the intervalley distance ($\delta k$) in the 1D BZ of a junction oriented along an angle $\theta$ relative to the zigzag crystallographic edge of the BLG sheet. Three situations are illustrated. The gray dashed lines show a general scenario with $0 < \theta < 30º$. The red/blue dashed lines correspond to $\theta = 0/30º$, i.e. when the junction is along the zigzag/armchair edge of the BLG sheet. **B**, AFM image of the K22 stack used to estimate the average orientation of the kink state junction. The white arrow points to a sharp edge of the BLG sheet. The yellow line marks the direction of the junction. The angle between them is $\theta = 6º$. **C**, The same for WK05. $\theta = 9º$. All scale bars are 2 μm.

4. **Device characterization**

**Table S1: The energy of ZA phonon in BLG**

| $\theta$ (°) | 0 | 6 | 9 | 21 | 24 | 30 |
|---|---|---|---|---|---|---|
| $\delta k$ (nm$^{-1}$) | 17 | 13.8 | 12.2 | 5.3 | 3.5 | 0 |
| Phonon energy (meV) | 68 | 54 | 47 | 11 | 6 | 0 |

We follow previous practices (*22*, *23*) to characterize the left and right top and bottom gates (LTG, LBG, RTG, RBG) and position the BLG in the two dual-gated regions at the charge neutrality point (CNP) with the desired *D*-field. As an example, Fig. S4 shows how we determine the gating relation for the RTG and RBG of device K22. The gating efficiency of the bottom gates is determined using the quantum Hall effect. Table S2 summarizes the parameters of each gate in K22. All gates are well behaved. The bulk *D*-field we can apply in our devices is limited by the breakdown of the weakest h-BN layer. In K22, this occurs between the bottom gates and the global gate at ~ 1.1 V/nm, which limits $|D_R| + |D_L|$ to roughly 500 mV/nm above 0.33 K. In WK05, we are able to reach $|D_R| = |D_L|$ ~ 450 mV/nm. These limits are not intrinsic to the device structure or the properties of the kink states.

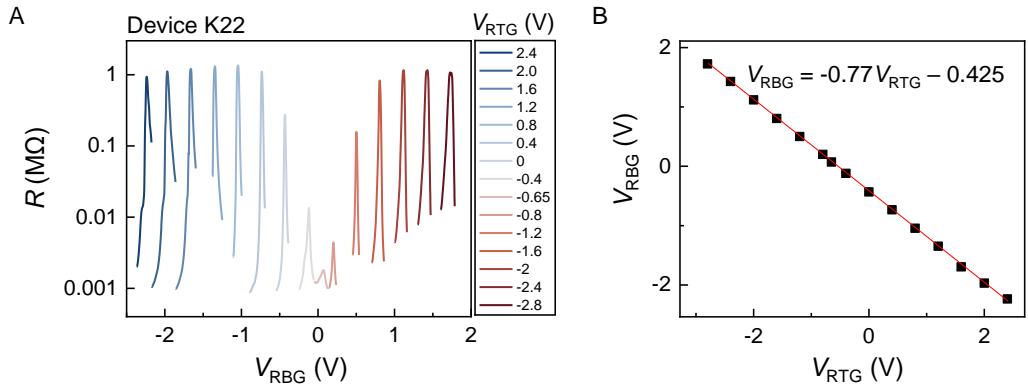

**Figure S4: Device characterization. A**, Junction resistance *R* as a function of $V_{RBG}$ at fixed $V_{RTG}$'s as indicated in the plot. This measurement allows us to obtain the gating relation of RBG and RTG and their respective $D = 0$ offsets. The results are shown in **B**. During measurement, the left region is positioned at the CNP with a large $D_L$ pointing in the same direction as $D_R$, i.e. $D_L > 0$ for $V_{RBG} > 0$ and vice versa, to minimize the left's contribution to conduction. *R* saturates at around 1 MΩ due to Joule heating of a constant current and parasitic capacitance of the cryostat wiring.

**Table S2: Gating characteristics in device K22**

| Gate | Gating efficiency ($10^{11}$ cm$^{-2}$ V$^{-1}$) | $D = 0$ voltage offset (V) | BN thickness (nm) |
|---|---|---|---|
| LTG/RTG | 4.72 | -0.67 | ~ 35 |
| LBG/RBG | 6.13 | 0.093 | ~ 27 |

## 5. Further evidence on the resistance quantization of the kink state

Figure S5 A-D shows the evolution of the junction resistance $R(V_{GG})$ in device K22 as the gating configuration changes from the non-topological regime, "− −", to the topological regime, "+ −" or "− +". In the "− −" regime, the CNP resistance increases with increasing $D$-field, reaching beyond MΩ at moderate $D$-field strength of ~ 200 mV/nm. In contrast, $R(V_{GG})$ exhibits a quantized resistance plateau very close to $h/4e^2$ in the "+ −" (Fig. S5C) or "− +" (Fig. S5D) gating configurations so long as the $D$-fields are not too small, i.e. > ~ 100 mV/nm. This corresponds to a bulk gap of ~ 10 meV in BLG (*39*, *53*).

In Figures 2B and S7E, we show exemplary $R(V_{GG})$ traces obtained in devices K24 and WK05 respectively, where the quantization of the kink state resistance is also very clear.

In Fig. 3 of the main text, we show that the kink state resistance remains constant in a wide temperature range above 0.3 K. In Fig. S6A here, we plot $R(V_{GG})$ taken at a few $D$-fields and $T$ = 20 m K, which requires loading the device into a different cryostat. The thermocycle changes the gate offsets and the resistance of the contacts slightly, resulting in a plateau value different from that of Fig. 3 by about 100 Ω. Nonetheless, it is clear that the quantization of the kink state resistance is a robust phenomenon insensitive to the temperature range of measurement. Further, after performing a linear fit to Eq. 1 and subtracting the contact resistance, we obtained $R_k = 6425$ Ω, in comparison to $R_k = 6443$ Ω obtained at 0.3 K (Table S3).

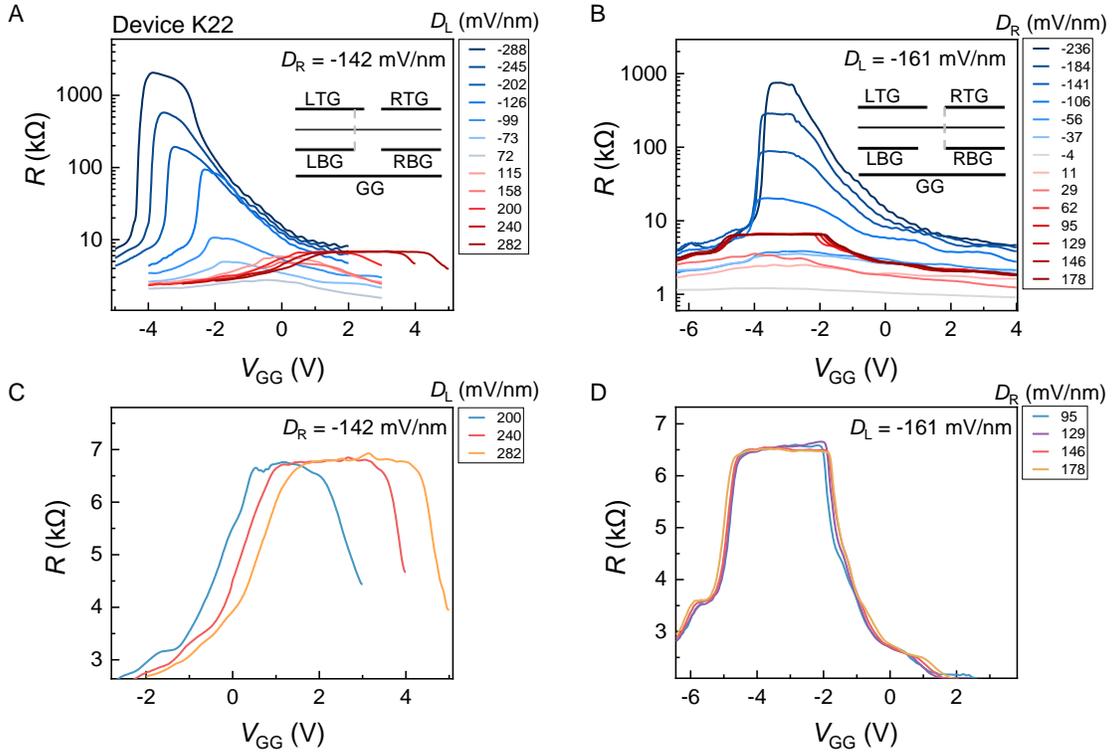

**Figure S5: The junction resistance in topological and non-topological gating configurations. A**, The junction resistance $R$ as a function of $V_{GG}$ with $D_R$ = -142 mV/nm and $D_L$ varying as labeled in the plot. **B**, $R(V_{GG})$ with $D_L$ = -161 mV/nm and $D_R$ varying as labeled. The resistance maximum of each trace corresponds to the CNP of the junction. The insets show the alignment of the gates, determined from the $V_{GG}$ shift of the CNP with changing $D_L/D_R$. **C** and **D** plot $R(V_{GG})$ of the topological regime in a linear scale for $D$-fields greater than ~ 100 mV/nm. From Device K22. $T$ = 1.9 K and $B$ = 0 T.

The kink states are expected to persist in the presence of a magnetic field. Our measurements confirm that this is indeed the case. Using measurements in Fig. 2C and similar traces, we plot in Fig. S6B the resistance value of the kink state plateau and the quantum Hall plateaus at n = 1 and 2 as a function of the magnetic field. All three decrease slightly with increasing $B$, likely due to a small change in the contact resistance. Overall, from 0 to 9 T, the kink state resistance changes only by 53 Ω (peak to peak), further attesting to its robust quantization.

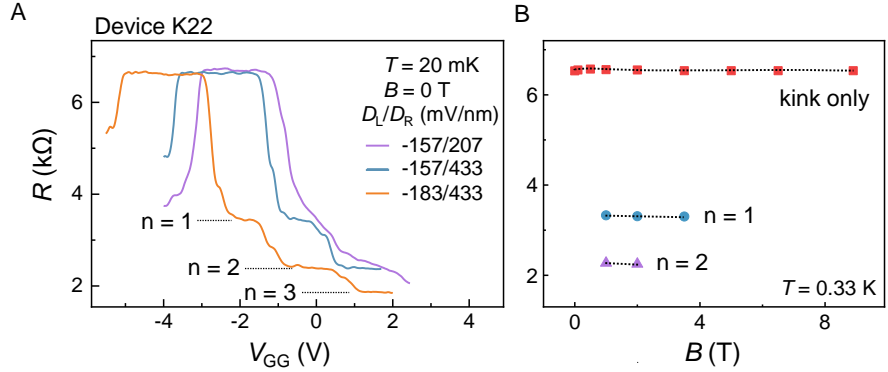

**Figure S6: A,** The junction resistance $R(V_{GG})$ measured at $T = 20$ mK for different $D$-fields as labeled in the plot. A linear fit to the kink state and the n = 1, 2, 3 plateaus of the orange trace yields $R_c = 225$ Ω and $R_k = 6425$ Ω. The n = 1, 2, 3 plateaus correspond to quantized conduction of the conventional 1D modes of the nanoribbon junction. They are fragile and only appear at very low temperatures. **B,** The $B$-field dependence of the resistance of the kink states, and the n = 1 and 2 quantum Hall states. $T = 0.33$ K. The kink state resistance varies by 53 Ω (peak to peak) from 0 T to 9 T. From 1 T to 3 T, the resistance of the kink state, the n = 1 and 2 plateaus decrease by 23 Ω, 26 Ω and 21 Ω, respectively. This is likely due to a small contact resistance change.

Table S3: The analysis of $R_c$ and $R_k$ in Device K22

| $B$ (T) | 0 | 0 | 1 | 2 |
|---|---|---|---|---|
| $T$ (K) | 0.02 | 0.33 | 0.33 | 0.33 |
| $D_L/D_R$ (mV/nm) | -183/433 | -161/146 | -161/146 | -161/146 |
| $R_{measure}$ (Ω) | 6650 ± 22 | 6557 ± 12 | 6557 ± 5 | 6548 ± 4 |
| $R_c$ (Ω) | 225 ± 13 | – | 114 ± 7 | 92 ± 5 |
| $R_k$ (Ω) | 6425 | – | 6443 | 6456 |
| Deviation of $R_k$ from $h/4e^2$ | 0.4% | – | 0.2% | < 0.1% |

## 6. Measurements on the temperature dependence of the junction resistance

Fig. S7 provides the raw $T$- and $D$-dependent $R(V_{GG})$ data obtained in Device WK05. The kink state plateau of WK05 has a slight tilt with $V_{GG}$, likely due to small contact resistance changes. To avoid selection bias, we average readings near the center within a $V_{GG}$ window as shown in the graphs and plot the results in Fig. 3B of the main text. Fig. S8 plots the value of $\Delta$ used in simulating the bulk resistance at different $D$-fields using $R_{bulk}^{-1} = R_1^{-1} e^{(-\Delta/2k_B T)} + R_2^{-1} e^{(-E_2/k_B T)}$ (Eq. 2 of the main text). Best fits to the high-temperature data yield values of $\Delta$ described by $\Delta$ [meV]= -5.7 + 0.13$D$ [mV/nm], in excellent agreement with our prior measurements of the band gap in BLG (*39, 53*). Other parameters used in the simulation are $R_1 = 1$ k$\Omega$, $R_2 = 200$ k$\Omega$, and $E_2 = 1.2$ meV, which are all consistent with previous findings (*39, 45, 53*). $R_1$ represents the band conduction in the high-$T$ limit. $R_2$ represents the efficiency of the hopping conduction, i.e. wave function overlap between neighboring sites and the number of impurity sites. $E_2$ is the average energy difference between neighboring sites $\langle \varepsilon_{ij} \rangle$. It is worth emphasizing that we have used the same set of $R_1, R_2$ and $E_2$ values to simulate $R_{bulk}^{-1}$ at *all* $D$-fields and are able to quantitatively capture *all* measurements. This agreement gives us strong confidence that a model of parallel conduction including the kink states and the bulk gap provides an excellent description of the measured junction resistance.

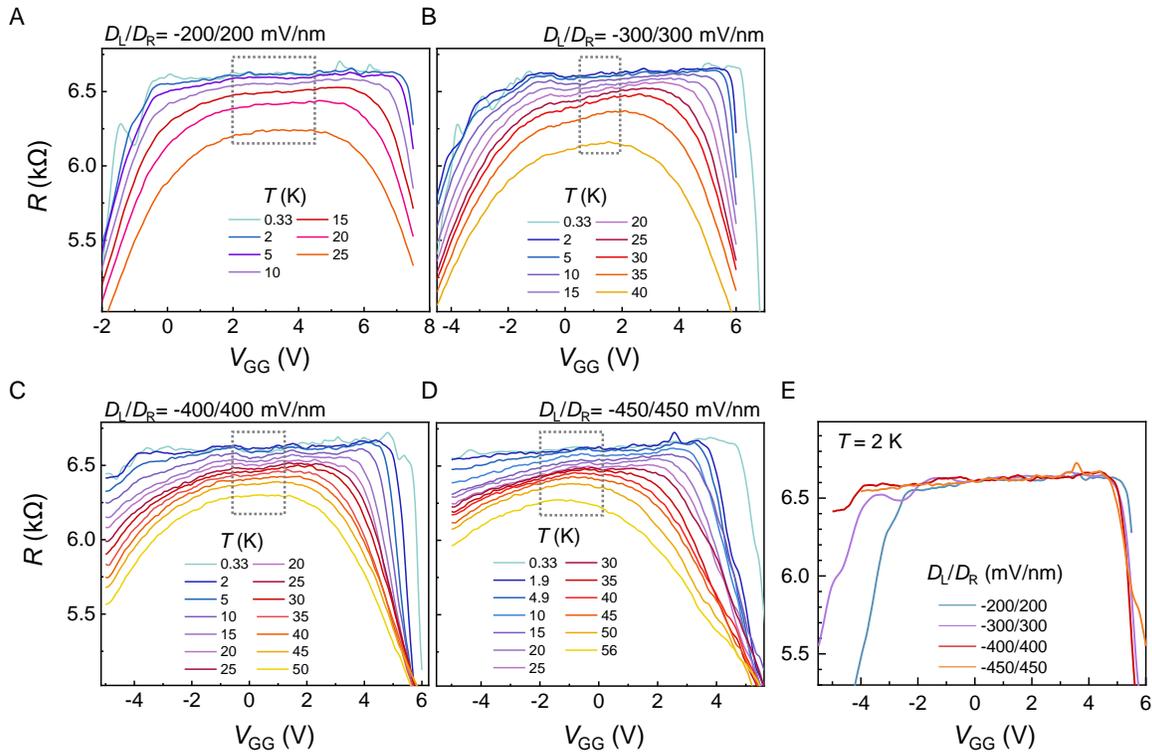

**Figure S7: The temperature and $D$-field dependence of the junction resistance in device WK05. A-D**, $R(V_{GG})$ measured at different temperatures and $D$-fields as labeled in the plot. $B = 0$ T. The data in the gray boxes are averaged for each temperature and plotted in Fig. 3B. **E**, Comparison of $R(V_{GG})$ taken at different $D$-fields as labeled. $T = 2$ K. The 200 mV/nm and 450 mV/nm traces are respectively shifted to the left/right by 2 V/1 V in $V_{GG}$ such that all four traces overlap at the right corner of the kink state plateau.

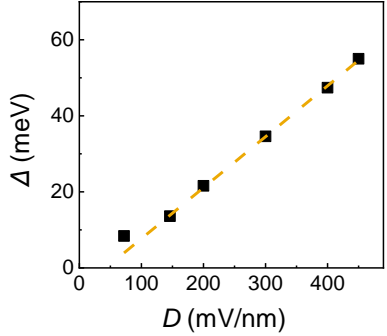

**Figure S8: $\Delta$ used in the simulation of $R_{bulk}$ for $D$ values used in measurements.** The orange dashed line is described by $\Delta$ [meV] = -5.7 +0.13$D$ [mV/nm], which is in excellent agreement with the $D$-dependence of the BLG gap obtained in Refs. (*39,53*).

## 7. Dynamic switching of the kink states

Much of the results reported here are obtained with steady-state gate voltages. In this section, we demonstrate electrically controlled dynamic switching of the kink states, which is technologically important. Figure S9 plots the measured junction conductance $G$ as the left gates are modulated to create a time-dependent, square waveform for $D_L$. The right gates are fixed to set $D_R = -170$ mV/nm. In half of the cycle, $D_L = -75$ mV/nm and $D_R = -170$ mV/nm. This places the junction in the "off" state and we observe a baseline conductance of $G = 0$. In the other half of the cycle, $D_L$ steps from 50 mV/nm to -75 mV/nm linearly along the black dashed line in the upper panel of Fig.S9. Similar to steady-state measurements, the junction turns on with $G = 4e^2/h$ in the "+ –" configuration and remains off in the "– –" configuration. The junction conductance $G$ saturates towards either $4e^2/h$ or 0 when the magnitude of $D_L$ reaches a few tens of mV/nm, corresponding to a small voltage swing of order 1V on the gates.

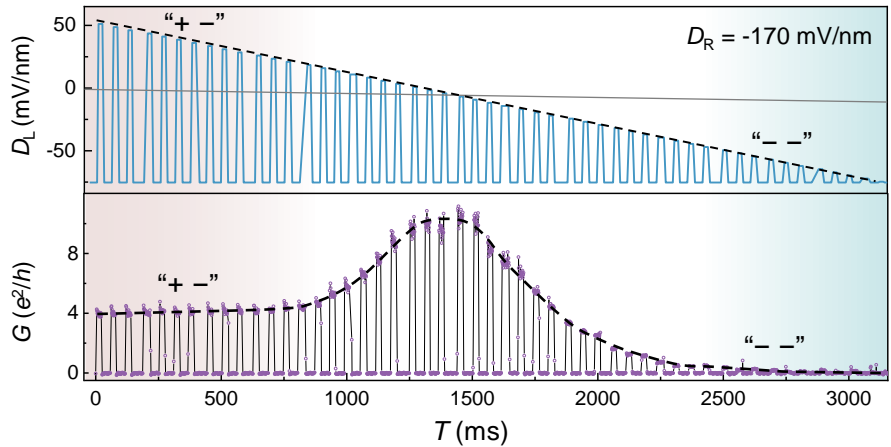

**Figure S9: Kink state as a dynamic switch.** The upper panel plots the time-dependent profile of $D_L$. The bottom panel plots the response of the junction conductance $G$ in units of $e^2/h$. The envelope of $G$ evolves with the envelope of $D_L$, similar to static measurements shown in Fig. 1F of the main text. The rise time of $G$ is approximately 6 ms, which is limited by how fast we can charge the gates. Data from K24.